\documentclass[12pt,onecolumn,amsfonts]{revtex4-1}
\usepackage{amsfonts}
\usepackage{amsmath}
\usepackage{amssymb}
\usepackage{graphics,graphicx}
\usepackage{psfrag}
\usepackage{epsf}
\usepackage{hyperref}
\usepackage[all]{hypcap}

\newcommand{\be}{\begin{equation}}
\newcommand{\ee}{\end{equation}}
\newcommand{\bea}{\begin{eqnarray}}
\newcommand{\eea}{\end{eqnarray}}
\newcommand{\bs}{\begin{equation}\begin{split}}
\newcommand{\es}{\end{split}\end{equation}}
\newcommand{\bal}{\begin{align}}
\newcommand{\eal}{\end{align}}

\begin{document}

\author{H. Landa$^1$, M. Drewsen$^2$, B. Reznik$^1$ and A. Retzker$^{3,4}$}
\title{Classical and Quantum Modes of Coupled Mathieu Equations}
\affiliation{$^1$School of Physics and Astronomy, Raymond and Beverly Sackler Faculty of Exact Sciences, Tel-Aviv University, Tel-Aviv 69978, Israel.
\\$^2$QUANTOP, Danish National Research Foundation Center for Quantum Optics, Department of Physics and Astronomy,University of Aarhus, DK-8000 \r{A}rhus C., Denmark.
\\$^3$Institut f{\"u}r Theoretische Physik, Universit{\"a}t Ulm, D-89069 Ulm.
\\$^4$Racah Institute of Physics, The Hebrew University of Jerusalem, Jerusalem 91904, Israel}

\begin{abstract}

We expand the solutions of linearly coupled Mathieu equations in terms of infinite continued matrix inversions, and use it to find the modes which diagonalize the dynamical problem. This allows obtaining explicitly the (`Floquet-Lyapunov') transformation to coordinates in which the motion is that of decoupled linear oscillators. We use this transformation to solve the Heisenberg equations of the corresponding quantum-mechanical problem, and find the quantum wavefunctions for stable oscillations, expressed in configuration-space. The obtained transformation and quantum solutions can be applied to more general linear systems with periodic coefficients (coupled Hill equations, periodically driven parametric oscillators), and to nonlinear systems as a starting point for convenient perturbative treatment of the nonlinearity.

\end{abstract}
\maketitle

\section{Introduction}\label{Sec:Intro}

The Mathieu equation for a single degree of freedom is very well known \cite{McLachlan}. In this paper we discuss the coupled system of Mathieu equations
\be\ddot{\vec{u}} +\left[A -2Q\cos 2t\right]\vec{u} = 0,\label{Eq:eomuHomogenousVec} \ee 
where $\vec{u}$ is an $f$-component vector and $A$ and $Q$ are constant symmetric $f\times f$ matrices. We also extend the treatment of eq.~\eqref{Eq:eomuHomogenousVec} to include an inhomogeneous right-hand-side, and more general $\pi$-periodic coupled parametric oscillators.

This multidimensional matrix equation has been researched thoroughly (see e.g.~\cite{Yakubovich} and \cite{Nayfeh} where also some applications are exemplified, and \cite{Takahashi,Hansen1985,Sinha}). Many general treatments of this system are perturbative and concerned with stability analysis - i.e. with finding the regions in some parameter-space for which the solutions are bounded. In this contribution we are interested primarily with describing the classical and quantum solutions of the system in terms of decoupled modes of oscillation. The solutions presented here may find application in the description of various discrete systems of coupled parametric oscillators, e.g., among others, trapped ion crystals \cite{major2005charged,werth2009charged}, coupled arrays of nanoelectromechanical oscillators \cite{LifshitzNonlinearReview,LifshitzNonlinearReview2} and binary Bose-Einstein condensates \cite{BinaryBEC}.

The quantum problem of time-dependent linear and quadratic Hamiltonians has also been considered in many publications. One of the first treatments is by Husimi \cite{Husimi1953}, who considered the problem of the one-dimensional quantum parametric oscillator with a driving force. He constructed Gaussian wavefunctions assuming that the classical solutions are known, and obtained the propagator of the system. He also derived transition amplitudes between states of Hamiltonians which are time-independent at some initial and final times.

The one-dimensional problem has been especially important to the description of the motion of single ions trapped in radiofrequency traps. It has been treated extensively using different methods and summarized in a few texts, e.g. \cite{WinelandReview,Schleich} and see references within.

In \cite{LewisReiesenfeld} Lewis and Reiesenfeld have considered a general time-dependent Hamiltonian, and have shown that the eigenvalues of an invariant operator (whose total time-derivative vanishes), are time-\textit{independent}, and that the eigenstates can be chosen with a specific time-dependent phase, so as to solve the Schr\"{o}dinger equation. This theory is the basis for most treatments of multidimensional time-dependent Hamiltonians.

In \cite{MalkinMankoTrifonovPhysRevD} coherent states and eigenfunctions have been obtained for a diagonal system of harmonic oscillators with a time-dependent frequency. Holz \cite{Holz1970} has considered multidimensional time-dependent oscillator Hamiltonians which remain positive-definite. He has constructed the coherent states, assuming that the Lewis-Riesenfeld invariants are known and that the Hamiltonian at the initial time is time-independent. Transition amplitudes have also been calculated in these works. In \cite{MalkinMankoTrifonov1973} expressions for the wavefunctions of general time-dependent upto-quadratic Hamiltonians are given formally using solutions of the classical equations in phase-space. Leach \cite{Leach1977} has considered the use of a time-dependent transformation to obtain a time-independent Hamiltonian, for the case of positive-definite Hamiltonians, and in terms of formal classical solutions. In this contribution we give the coherent and number states of the Schr\"{o}dinger equation explicitly in terms of the decoupled modes of eq.~\eqref{Eq:eomuHomogenousVec}.

This paper is organized as follows. In Sec.~\ref{Sec:Solution} we obtain an analytic expansion for the solutions of eq.~\eqref{Eq:eomuHomogenousVec}, which is not based on a small parameter, but rather uses infinite continued matrix inversions. We obtain explicitly a time-dependent transformation to coordinates in which the motion is that of decoupled linear oscillators. In Sec.~\ref{Sec:Quantization} we use this transformation to find the wavefunctions of the corresponding quantum system. We conclude in Sec.~\ref{Sec:Conclusion} with a summary of our results and comments of possible further research and applications to various physical systems. The classical linearized modes of an ion crystal in a Paul trap are treated in \cite{rfions} using the methods described here.

\section{Solution of the Linear Equations}\label{Sec:Solution}

\subsection{The Floquet Problem}\label{Sec:Floquet}

Eq.~\eqref{Eq:eomuHomogenousVec} is a homogenous linear differential equation with periodic coefficients and therefore amenable to treatment using Floquet theory. In this subsection we recall a few facts from this theory and introduce the Floquet-Lyapunov transformation which will allow us to obtain explicitly the classical and quantum solutions.

Eq.~\eqref{Eq:eomuHomogenousVec}, with a $\pi$-periodic parametric drive, can be obtained by a suitable nondimensionalization (scaling) of the coordinates and time of the equations of motion (e.o.m) of a physical system parametrically driven at frequency $\Omega$. For the Newtonian problem with $f$ degrees of freedom, the corresponding Floquet problem is stated in terms of coordinates in $2f$-dimensional phase space by defining
\be\phi =\left(\begin{array}{c} {\vec{u}} \\ {\dot{\vec{u}}} \end{array}\right),\qquad \Pi \left(t\right)=\left(\begin{array}{cc} {0} & {1_{f} } \\ {-\left(A-2Q\cos 2t\right)} & {0} \end{array}\right) \label{Eq:Pioft},\ee
where $1_{f} $ is the $f$-dimensional identity matrix. The e.o.m is written in standard form as
\be \dot{\phi }=\Pi \left(t\right)\phi \label{Eq:eomFloquet}. \ee 

In the following, an $f$-dimensional vector $\vec{u}$ will be denoted by a lower case Latin letter (usually with an arrow), and Latin subscripts will be used for its components ($u_{m} $). $f$-dimensional matrices will be denoted by capital Latin letters ($Q$). A $2f$-dimensional vector $\phi $ will be denoted by a lower case Greek letter (with no arrow), and Greek subscripts will be used for components of $2f$-dimensional vectors ($\phi _{\nu } $). Capital Greek letters (unitalicized) will denote $2f$-dimensional matrices ($\Pi $, or {\rm B}).

A fundamental matrix solution to eq.~\eqref{Eq:eomFloquet} has $2f$ linearly independent column solutions and obeys the matrix equation
\be\dot{\Phi }\left(t\right)=\Pi \left(t\right)\Phi \left(t\right)\label{Eq:eomFloquetPhi}.\ee
 A principal fundamental matrix solution $\Phi \left(t\right)$ is a fundamental matrix solution which equals the identity matrix at some point in time. We will always take a principal fundamental matrix solution to obey this at $t=0$, i.e.
\be\Phi \left(0\right)=1_{2f} \label{Eq:Phit0}.\ee
and then $\Phi$, which is obviously unique, is also known as the \textit{matrizant} of eq.~\eqref{Eq:eomFloquet}). 

Now let $T=2\pi /\Omega =\pi $ be the period of $\Pi \left(t\right)$, i.e. $\Pi \left(t+T\right)=\Pi \left(t\right)$. Then $\Phi \left(t+T\right)$ is also a fundamental matrix solution. Therefore its columns must be linear combinations of the columns of $\Phi \left(t\right)$, i.e. $\Phi \left(t+T\right)=\Phi \left(t\right)\Xi $ for some non-singular constant matrix $\Xi$.
In particular, given the initial conditions in eq.~\eqref{Eq:Phit0}, we find that $\Xi =\Phi \left(T\right)$, known as the \textit{monodromy matrix}. $\Xi $ can be brought to Jordan canonical form, which we assume to be diagonal (which holds for the case of stable oscillations). The diagonalization is given by
\be {\rm P} ^{-1} \Xi {\rm P} = \Lambda ={\rm diag}\left\{\lambda _{1} ,...,\lambda _{2f} \right\} \ee
where $\lambda _{\nu}$ are the complex Floquet eigenvalues (also known as multipliers).

Applying the coordinate transformation by writing 
\be\Phi \left(t\right)=\Upsilon \left(t\right) {\rm P} ^{-1}\label{Eq:PhiCoordChange}\ee
we get that $\Upsilon \left(t+T\right)=\Upsilon \left(t\right) {\rm P} ^{-1} \Xi {\rm P} =\Upsilon \left(t\right) \Lambda $. Therefore the $\nu $-th column-vector of $\Upsilon \left(t\right)$ obeys $\Upsilon _{\nu } \left(t+T\right)=\lambda _{\nu } \Upsilon _{\nu } \left(t\right)$, and thus $\Upsilon _{\nu } \left(t+nT\right)=\lambda _{\nu } ^{n} \Upsilon _{\nu } \left(t\right)$. 
Consequently the solutions are decaying for $\left|\lambda _{\nu } \right|<1$ and unstable for $\left|\lambda _{\nu } \right|>1$. When eqs.~\eqref{Eq:Pioft}-\eqref{Eq:eomFloquet} are Hamiltonian (as in our case), the sets $\left\{\lambda_{\nu}\right\}$ and $\left\{\left(\lambda_{\nu}^*\right)^{-1}\right\}$ must coincide. Since the equations have real coefficients, nonreal $\lambda _{\nu }$ must come in conjugate pairs. Therefore nonreal $\lambda _{\nu }$ not on the unit circle come in quadruplets as $\lambda _{\nu }$, $\left(\lambda_{\nu}^*\right)^{-1}$, $\lambda_{\nu }^*$, $\lambda _{\nu }^{-1}$ and real $\lambda _{\nu }$ not on the unit circle come in pairs as  $\lambda _{\nu }$, $\lambda _{\nu }^{-1}$.
Letting $\lambda _{\nu } =e^{i\beta _{\nu } T} $ defines the characteristic exponents $\beta _{\nu } =\frac{1}{iT} \ln \lambda _{\nu }$. 
We multiply $\Upsilon _{\nu } \left(t+T\right)=\lambda _{\nu } \Upsilon _{\nu } \left(t\right)$ by $e^{-i\beta _{\nu } \left(t+T\right)} $ and obtain the normal form for the column-vector solutions 
\be\Upsilon _{\nu } \left(t\right)=\Gamma _{\nu } \left(t\right)e^{i\beta _{\nu } t}\label{Eq:Upsilon} ,\ee where $\Gamma _{\nu }$ are $T$-periodic vectors, $\Gamma _{\nu } \left(t+T\right)=\Gamma _{\nu } \left(t\right)$.

Using eqs.~\eqref{Eq:PhiCoordChange} and \eqref{Eq:Upsilon} and defining
\be{\rm B} ={\rm diag}\left\{i\beta _{1} ,...,i\beta _{2f} \right\}, \label{Eq:B}\ee
 we may now write
\be\Phi \left(t\right) =\Gamma \left(t\right)e^{{\rm B} t} {\rm P}^{-1} = \Gamma \left(t\right)e^{{\rm B} t} \Gamma ^{-1}\left(0\right) \label{Eq:Phit}.\ee 
We note that in the general case, ${\rm P}^{-1}=\Gamma ^{-1}\left(0\right)$ will not be unitary, although this may happen in certain highly symmetric cases, e.g., in the trivial case in which $A$ and $Q$ commute and there exists a constant orthogonal transformation which diagonalizes eq.~\eqref{Eq:eomuHomogenousVec} into a system of decoupled Mathieu equations.

Differentiating eq.~\eqref{Eq:Phit} and substituting into the e.o.m, eq.~\eqref{Eq:eomFloquetPhi}, we have
\be \left(\dot{\Gamma }e^{{\rm B} t} +\Gamma Be^{{\rm B} t} \right) {\rm P}^{-1}=\Pi \Gamma e^{{\rm B} t}{\rm P}^{-1} \ee
or,
\be \dot{\Gamma}+\Gamma {\rm B} =\Pi \Gamma. \label{Eq:PZB}\ee 
If we now substitute into eq.~\eqref{Eq:eomFloquet} the time-dependent coordinate change 
\be \phi \left(t\right)=\Gamma\left(t\right)\chi \left(t\right)\label{Eq:FLTrans},\ee
 we get 
\be\dot{\Gamma}\chi + \Gamma \dot{\chi }=\Pi \Gamma\chi, \ee 
which, after using eq.~\eqref{Eq:PZB} and multiplying on the left by $\Gamma^{-1} $, reduces to
\be\dot{\chi }={\rm B} \chi .\label{Eq:eomchiLinear}\ee 

Eq.~\eqref{Eq:eomchiLinear} is a diagonal equation with constant coefficients, the solutions of which are the Floquet modes
\be \chi _{\nu } \left(t\right)=\chi _{\nu } \left(0\right)e^{i\beta _{\nu } t}. \ee
The time-dependent transformation $\Gamma \left(t\right)$ of eq.~\eqref{Eq:FLTrans} is known as the Floquet-Lyapunov transformation.
 
\subsection{Solution Using an Expansion in Infinite Continued Matrix Inversions}\label{Sec:MathieuSolution}

We turn to an analytical expansion of the solutions of the homogenous e.o.m \eqref{Eq:eomuHomogenousVec}.
The following expansion allows to obtain the frequencies and the coefficients of the solution vectors in a generalization of an infinite continued fractions expansion, to arbitrary precision. Infinite recurrence relations have been used for solving various types of differential equations (see e.g. chapter 9 of \cite{Risken}) and differential-delay equations \cite{Pelster1999}, and applied recently to the study of the stability of a trapped Bose-Einstein condensate \cite{PelsterBEC}. The method described below gives the solution in a form which is immediately suitable for obtaining the Floquet-Lyapunov transformation. 

We seek for solutions of eq.~\eqref{Eq:eomuHomogenousVec}, in the form of a sum of two linearly independent complex solutions,
\be \vec{u} =\sum _{n=-\infty }^{n=\infty }\vec{C}_{2n} \left[b e^{i\left(2n+\beta \right)t} +c e^{-i\left(2n+\beta \right)t} \right] ,\label{Eq:uAnsatz}\ee 
where $b$ and $c$ are complex constants determined by the initial conditions. In general, the characteristic exponents $\beta$ may be complex. Except when there are $\beta$'s which are real and integral, a system composed of solutions of the form of eq.~\eqref{Eq:uAnsatz} will constitute a fundamental system. For an integral $\beta$, eq.~\eqref{Eq:uAnsatz} gives a single $\pi$- or $2\pi$-periodic solution and the other linearly independent solution will in general (excluding the trivial case $Q=0$) be unbound \cite{McLachlan,Yakubovich}. We do not treat this marginal case as we will soon restrict ourselves to stable oscillations only. Following Sec.~\ref{Sec:Floquet}, stable modes will be described by $\beta$ taking a real nonintegral value, which can obviously be chosen in the range $0<\beta<2,\beta\neq 1$. For the stable modes, when $\beta$ is real, we find that $b=c^*$ and $\vec{C}_{2n}$ are all real.

We assign eq.~\eqref{Eq:uAnsatz} in eq.~\eqref{Eq:eomuHomogenousVec}, discard the negative exponent terms (which give identical relations), and find that the solutions must obey for all $t$
\be-\sum\vec{C}_{2n} \left(2n+\beta \right)^{2} e^{i\left(2n+\beta \right)t} +\left[A -Q \left(e^{i2t} +e^{-i2t} \right)\right]\sum\vec{C}_{2n} e^{i\left(2n+\beta \right)t} =0,\ee
where in the above expression and for the rest of this section, the summation is over $n \in {\mathbb Z}$. Thus we get the recursion relation
\be-\vec{C}_{2n} \left(2n+\beta \right)^{2} +A \vec{C}_{2n} -Q \left(\vec{C}_{2n-2} +\vec{C}_{2n+2} \right) =0.\ee 
By defining
\be R_{2n} =A-\left(2n+\beta \right)^{2},\ee 
we can write the first infinite recursion relation, which progresses towards positive values of $n$,
\be Q\vec{C}_{2n-2} =R_{2n} \vec{C}_{2n} - Q\vec{C}_{2n+2}. \label{Eq:Recursion1}\ee 
To get a second relation which progresses towards negative values of $n$, we reorganize eq.~\eqref{Eq:Recursion1}, obtaining 
\be Q\vec{C}_{2n+2} = R_{2n} \vec{C}_{2n} -Q\vec{C}_{2n-2}.\label{Eq:Recursion2}\ee 

From these relations we can now obtain an expansion in infinite continued matrix inversions. Starting with $n=1$ in eq. \eqref{Eq:Recursion1} we get by repeated substitutions 
%\[C_{0} =Q^{-1} R_{2} C_{2} -C_{4} \] \[C_{2} =Q^{-1} R_{4} C_{4} -C_{6} \] \[C_{4} =Q^{-1} R_{6} C_{6} -C_{8} \] \[C_{6} =Q^{-1} R_{8} C_{8} -C_{10} \] 
%\bm \vec{C}_{2} =T_{2,\beta } \vec{C}_{0} \equiv \\ \left[Q^{-1} R_{2} -\left[Q^{-1} R_{4} -\left[Q^{-1} R_{6} -... \right]^{-1} \right]^{-1} \right]^{-1} \vec{C}_{0}. \label{Eq:MatrixInversions1} \end{multline}
\be \vec{C}_{2} =T_{2,\beta }Q \vec{C}_{0} \equiv  \left(\left[ R_{2} -Q\left[ R_{4} - Q\left[ R_{6} -... \right]^{-1}Q \right]^{-1}Q \right]^{-1}\right)Q \vec{C}_{0}. \label{Eq:MatrixInversions1} \ee
Substituting decreasing values of $n$ starting with $n=0$ in eq. \eqref{Eq:Recursion2}, we get the independent relation
\be Q\vec{C}_{2} = R_{0} \vec{C}_{0} -Q\vec{C}_{-2} =\tilde{T}_{0,\beta }\vec{C}_{0} \equiv \left(R_{0} -Q\left[ R_{-2} -Q\left[ R_{-4} -... \right]^{-1}Q \right]^{-1}Q \right)\vec{C}_{0}. \label{Eq:MatrixInversions2} \ee 

Multiplying eq.~\eqref{Eq:MatrixInversions1} by $Q$ and defining
\be Y_{2,\beta } \equiv \tilde{T}_{0,\beta } - Q T_{2,\beta } Q\label{Eq:Y2beta}, \ee
we find that all characteristic exponents $\beta $ are zeros of the determinant of $Y_{2,\beta } $ (which is a function of $\beta $). If there are degenerate $\beta $'s they will appear as degenerate zeros of this determinant. The vector $\vec{C}_{0} $ for each $\beta $ is an eigenvector of $Y_{2,\beta } $ with eigenvalue $0$. Since $A$ and $Q$ are symmetric, $Y_{2,\beta } $ is symmetric as well, and so its kernel will be of dimension equal to the algebraic multiplicity of the $\beta$ root. The vector $\vec{C}_{2} $ can be obtained by an application of $T_{2,\beta }Q $ to $\vec{C}_{0} $, for $n=-1$ we use $\vec{C}_{-2} = \left[T_{-2,\beta }\right]^{-1}Q \vec{C}_{0}$, and so on for the other vectors. We note that the different vectors $\vec{C}_{2n,\beta } $ are not orthogonal in general, and the vectors at every order in $n$ mix different coordinates.

Because of the presence of the diagonal term $\left(2n+\beta \right)^{2} $ in $R_{2n} $, we would have $\left\| R_{2n} \right\| _{1} \propto \left(2n+\beta \right)^{2} +{\rm O} \left(1\right)$, and therefore the general term of the expansion vanishes. Either $A$ or $Q$ may be singular and the expansion can still be applied in general. Even if both are singular, the expansion is valid if there are no integral values of $\beta$, a case which we do not tackle as noted above. Excluding perhaps isolated values of $\beta$ (and atypically in the parameter-space), all matrices which are inverted in the above expressions will be invertible, and while employing the algorithm in practice, the invertibility of the matrices is of course easily verified at each step. In App.~\ref{App:Inversions} we extend the infinite matrix inversions to obtain the periodic solution of eq.~\eqref{Eq:eomuHomogenousVec} with an inhomogenous r.h.s, and also comment on some computational aspects of this method. In App.~\ref{App:DoubleCos} we show briefly how the method may be extended to a system of coupled Hill equations.

%For the unstable modes, $\beta$ is imaginary and the realness of $u_m$ gives the further condition \be \vec{C}_{-2n} = \vec{C}_{2n}^*.\ee

\subsection{The Floquet-Lyapunov Transformation For Stable Modes}\label{Sec:FLTrans}

 We can now find explicitly the time-dependent Floquet-Lyapunov transformation of eq.~\eqref{Eq:FLTrans} which transforms the Floquet problem to a time-independent equation, which in our construction is also diagonal. We further assume that all Floquet modes are stable, i.e. that the $2f$ linearly-independent solutions of eq.~\eqref{Eq:eomFloquet} are oscillatory and thus characteristic exponent come in complex conjugate pairs. This simplifies many expressions and avoids complications in the quantization, since the eigenfunctions of the negative harmonic potential (the parabola potential \cite{harmonic_parabola}) are not square integrable over the real line.
We therefore take ${\rm B}$ of eq. \eqref{Eq:B} in the block form
\be {\rm B}= \left(\begin{array}{cc} {iB} & { 0} \\ {0 } & {-iB }\end{array}\right), \qquad \left(B\right)_{f\times f} = {\rm diag}\left\{\beta_1,...,\beta_f\right\} \label{Eq:Bblock}\ee
where $\beta_j$ are positive. We define the $f$-dimensional matrix $U$ whose columns are constructed from the series of $f$-dimensional vectors $\vec{C}_{2n,\beta _{j}}$ obtained from the recursion relations for the solutions $\vec{u}$ of eq.~\eqref{Eq:uAnsatz}, i.e.
\be \left(U\right)_{f\times f} = \left(\begin{array}{cc} 
{\sum \vec{C}_{2n,{\beta _{j} }} e^{i2nt} } & { ... } \end{array}\right)\ee
 We similarly define the $f$-dimensional matrix $V$ composed from column-vectors as
\be \left(V\right)_{f\times f} = \left(\begin{array}{cc} 
{i\sum \left(2n+\beta _{j} \right)\vec{C}_{2n,\beta _{j}} e^{i2nt}} & { ... } \end{array}\right), \ee
and thus we may represent a $2f$-dimensional fundamental matrix solution in the form $\Psi \left(t\right)e^{{\rm B}t}$ where $\Psi \left(t\right)$ is written in block form 
\be\Psi \left(t\right)=\left(\begin{array}{cc} {U} & { U^*} \\ {V } & {V^* }\end{array}\right),\ee 
where $U^*$ denotes the complex conjugate (and not hermitian conjugate) of the matrix $U$.

%\begin{widetext} \be\Psi \left(t\right)=\left(\begin{array}{cccc} {\sum \vec{C}_{2n,{\beta _{j} }} e^{i2nt}  } & {...} & {\sum\vec{C}_{2n,{\beta _{j} }} e^{-i2nt}  } & {...} \\ {i\sum \left(2n+\beta _{j} \right)\vec{C}_{2n,\beta _{j}} e^{i2nt}  } & {...} & {-i\sum \left(2n+\beta _{j} \right)\vec{C}_{2n,\beta _{j}} e^{-i2nt}  } & {...} \end{array}\right)\ee \end{widetext}
 By multiplying $\Psi \left(t\right)e^{{\rm B}t}$  on the right with $\Psi ^{-1} \left(0\right)$ we get the matrizant $\Phi\left(t\right)$ since the initial condition of eq.~\eqref{Eq:Phit0} is obeyed. Then by comparing with eq.~\eqref{Eq:Phit} we find
\be \Psi \left(t\right)e^{{\rm B}t}\Psi ^{-1} \left(0\right)=\Gamma \left(t\right) e^{{\rm B}t}\Gamma ^{-1} \left(0\right),\ee
so that we may \textit{choose} 
\be\Gamma \left(t\right)=\Psi \left(t\right)=\left(\begin{array}{cc} {U} & { U^*} \\ {V } & {V^* }\end{array}\right)\label{Eq:Gamma}\ee
 (the choice is in fact unique only up to a constant matrix which commutes with ${\rm B}$, and we will use this fact in the following). %using the identity $e^{C^{-1}AC}C^{-1}=C^{-1}e^A $ and 

\section{Quantization}\label{Sec:Quantization}

\subsection{Hamiltonian Formalism}\label{Sec:Hamiltonian}

In this subsection we consider the results of the previous section within the Hamiltonian formalism. We find the conditions for the Floquet-Lyapunov transformation to be canonical, which also allows us to obtain its inverse explicitly in terms of matrix transpositions and complex conjugation. This requires finding the generating function of classical mechanics, and gives the transformed Hamiltonian. These results are novel to the best of our knowledge.

Let us rewrite the e.o.m, eq.~\eqref{Eq:eomFloquet}, in the form
\be{\rm J}\dot{\phi }={\rm J}\Pi \phi \equiv {\rm H} \phi, \label{Eq:eomCanonical}\ee
where
\[{\rm J}=\left(\begin{array}{cc} {0} & { -1_f} \\ { 1_f } & {0}\end{array}\right),\qquad {\rm H} = \left(\begin{array}{cc} {A -2Q\cos 2t} & {0} \\ { 0} & {1_f }\end{array}\right).\]
 ${\rm J}$ is a skew-symmetric matrix (${\rm J}^{-1}={\rm J}^t=-{\rm J}$) and ${\rm H}$ is symmetric. Denoting by $\vec{p}=\dot{\vec{u}}$ the momenta canonically conjugate to the coordinates $\vec{u}$, eq.~\eqref{Eq:eomCanonical} is seen to be the canonical form of Hamilton's equation,
\be \dot{u}_j = \frac{\partial{\mathcal H}}{\partial p_j}, \qquad \dot{p}_j = -\frac{\partial{\mathcal H}}{\partial u_j},\ee
with the corresponding Hamiltonian ${\mathcal H}$ written as the quadratic form 
\be {\mathcal H} = \frac{1}{2} \phi^t {\rm H} \phi, \label{Eq:Hphi} \ee
where $\phi^t$ denotes the transposed vector. 
%, and which can be used to show that for any two solutions $\phi\left(t\right)$ and $\varphi\left(t\right)$ of eq.~\eqref{Eq:eomCanonical}, \be \phi^\dag\left(t\right){\rm J}\varphi\left(t\right)={\rm const}. \ee

Similarly, we rewrite the transformed e.o.m, eq.~\eqref{Eq:eomchiLinear}, in the form
 \be{\rm K}\dot{\chi }= {\rm K B}\chi, \label{Eq:eomHamiltonian}\ee
 where ${\rm K}$ is the antihermitian matrix (${\rm K}^{-1}={\rm K}^\dag=-{\rm K}$)
\[{\rm K}=\left(\begin{array}{cc} {-i 1_f} & { 0} \\ {0 } & {i 1_f}\end{array}\right),\]
and ${\rm KB}$ is a hermitian matrix (in fact, positive definite), with the explicit form ${\rm KB}= {\rm diag}\left\{\beta _{1} ,...,\beta _{f},\beta _{1},...,\beta _{f} \right\}$.

Let us here introduce explicitly the connonically conjugate variables of the Hamiltonian formalism (we here break the notation a little),
\[ \chi = \left(\begin{array}{c} {\xi} \\ {\zeta} \end{array}\right),\]
where $\xi$ is the new $f$-dimensional vector of coordinates and we see that $-i\zeta$ are the new conjugate momenta, such that eq.~\eqref{Eq:eomHamiltonian} is derivable from the Hamiltonian
\be {\mathcal H}' \equiv \frac{1}{2}\chi^t \tilde{\rm H} \chi + {\mathcal Y}\left(t\right), \label{Eq:Hamiltonian2} \ee
with
\be {\rm \tilde{H}} = \left(\begin{array}{cc} {0} & {B} \\ {B} & {0}\end{array}\right)\label{Eq:tildeH},\ee
$B$ is given by eq.~\eqref{Eq:Bblock} and ${\mathcal Y}\left(t\right)$ is a function of time alone, that does not enter the equations of motion. In the following we will prove that indeed ${\mathcal H}'$ is the transformed Hamiltonian \textit{provided} that the Floquet-Lyapunov transformation is canonical. Our choice of the Floquet-Lyapunov transformation will be such that classically, ${\mathcal Y}\left(t\right) =0$. Using the realness of $\vec{u}$ and $\vec{p}$ and the explicit expressions for $\Gamma\left(t\right)$ and $\Gamma^{-1}\left(t\right)$ it is easy to verify that
\be \xi = \zeta^*. \label{Eq:Realness} \ee

In App.~\ref{App:Hamiltonian} we obtain an expression for the inverse of the Floquet-Lyapunov transformation, and show that the matrices $U$ and $V$ can be rescaled by multiplication with a (diagonal) matrix, such that
\be U\to U\left(-2iV^t\left(0\right)U\left(0\right)\right)^{-\frac{1}{2}},\label{Eq:ScalingMatrix}\ee
and $V$ accordingly, thereby imposing the following normalization condition,
\be 
V^t\left(0\right)U\left(0\right) = \frac{1}{2}i.\label{Eq:Normalization}\ee
By eqs.~\eqref{Eq:AppGammaInv}-\eqref{Eq:M} subject to eq.~\eqref{Eq:Normalization}, we get
\be\Gamma^{-1} \left(t\right)=\left(\begin{array}{cc} {iV^\dag } & {-iU^\dag} \\ {-iV^t} & {iU^t}\end{array}\right),\label{Eq:GammaInv}\ee
and by writing $\Gamma^{-1}\Gamma=1_{2f}$ in block form, we find the identities
\be U^tV^*-V^tU^*=-i,\qquad U^tV=V^tU \label{Eq:Identities}\ee
which we will use below.

We now turn to finding the classical generating function of the canonical transformation relating ${\mathcal H}$ to ${\mathcal H}'$. If $\chi\left(t\right) = \Gamma^{-1}\left(t\right) \phi\left(t\right)$ is to be a canonical transformation, we search for the generating function ${\mathcal F}\left( \vec{u}, -i\zeta, t\right)$, expressed in terms of the old coordinates and new momenta, which obeys the set of equations
\be p_j = \frac{\partial {\mathcal F}}{\partial u_j}, \qquad \xi_j = \frac{\partial {\mathcal F}}{\partial\left(-i \zeta_j\right)}, \label{Eq:GeneratingConditions}\ee
and the transformed Hamiltonian is given by
\be{\mathcal H}' = {\mathcal H} + \partial {\mathcal F}/\partial t. \label{Eq:FormalTransformedH}\ee

With the help of eq.~\eqref{Eq:GammaInv} we can invert for $\vec{p}$ in terms of $\vec{u}$ and $\zeta$ to get
\[ \vec{p}=-iU^{-t}\zeta+U^{-t}V^{t}\vec{u},\qquad \xi=iV^{\dag}\vec{u}-iU^{\dag}\left(U^{-t}V^{t}\vec{u}-iU^{-t}\zeta\right),\]
 where we define for bravity $U^{-t} \equiv \left[U^{-1} \right]^{t}$, and in the following we will use $O ^{-\dag}$ and $O ^{-*}$ with a similar definition. A solution to eqs.~\eqref{Eq:GeneratingConditions} exists and reads
\be {\mathcal F}=\frac{1}{2} \vec{u}^t U^{-t}V^{t}\vec{u} -i\vec{u}^t U^{-t}\zeta + \frac{1}{2}i\zeta^tU^{\dag} U^{-t}\zeta \label{Eq:F} \ee
\textit{provided} that
\be V^{\dag} -U^{\dag}U^{-t}V^t=-i U^{-1} \label{Eq:CanonicalCondition}\ee
and that $U^{-t}V^{t}$ and $U^{\dag}U^{-t}$ are symmetric matrices. These conditions follow after some manipulations from the identities in eq.~\eqref{Eq:Identities}. Thus we see that the normalization of eq.~\eqref{Eq:Normalization} guarantees that the Floquet-Lyapunov transformation is canonical.

%It is easily seen that for the generating function ${\mathcal G}\left( \zeta, \vec{u},t\right)$ of the inverse transformation $\phi\left(t\right) = \Gamma\left(t\right) \chi\left(t\right)$, to have the affect of reverting the Hamiltonian ${\mathcal H}'$ back to ${\mathcal H}$, it must satisfy $-i{\mathcal F}+{\mathcal G}=0$, i.e. \be FIX {\mathcal G}=-\frac{1}{2}\vec{u}^t U^{-\dag}V^{\dag}\vec{u} - i\vec{u}^t U^{-\dag}\zeta + \frac{1}{2}i\zeta^tU^t U^{-\dag}\zeta \label{Eq:G}. \ee
%since it obeys the equations \be \tilde{\lambda} \xi_j = \partial {\mathcal G}/\partial \zeta_j, \qquad p_j = - \partial {\mathcal G}/\partial u_j, \label{Eq:GeneratingConditions2}\ee \be \tilde{\lambda} = \lambda^{-1}, {\mathcal G} = {\mathcal F}.\ee

\subsection{Quantization}

In this subsection we apply the Floquet-Lyapunov transformation to the operators in the Heisenberg picture of the quantum problem corresponding to eq.~\eqref{Eq:eomuHomogenousVec}, allowing to diagonalize it in terms of ladder operators. We then find explicit expressions for the wavefunctions of the coherent and number states, utilizing the periodicity of the Floquet-Lyapunov transformation and its inverse, and the decoupled oscillatory modes with their characteristic frequencies. These results are novel to the best of our knowledge.

We first canonically quantize the system by promoting the canonically conjugate variables $\vec{u}$ and $\vec{p}$ to operators obeying the quantum commutation relations
\be \left[\hat{p}_j,\hat{u}_k\right] = -i\delta_{jk}, \qquad \left[\hat{p}_j,\hat{p}_k\right] = \left[\hat{u}_j,\hat{u}_k\right] = 0, \label{Eq:CommutationRelations}\ee
%\left[\hat{\phi}_j,\hat{\phi}_{f+k}\right] = 
where we will denote operators with a hat, and set $\hbar = 1$. The Heisenberg equations of motion for these operators, $\hat{\phi}$, are identical to eq.~\eqref{Eq:eomCanonical}. Repeating the derivation of Sec.~\ref{Sec:Floquet} we see that the noncommutativity of the operators has no affect on the transformation and thus we find in the Heisenberg picture the e.o.m
\be{\rm K}\dot{\hat{\chi}}= \tilde{\rm H} \hat{\chi}, \label{Eq:eomOperators}\ee 
which, using eq.~\eqref{Eq:Bblock}, is the diagonal set
\be \dot{\hat{\xi}}_j= i\beta_j\hat{\xi}, \qquad \dot{\hat{\zeta}}_j= -i\beta_j\hat{\zeta}_j,\ee 
the solution of which is simply
\be\hat{\xi}_j\left(t\right)= \hat{\xi}_j\left(0\right)e^{i\beta_jt}, \qquad
\hat{\zeta}_j\left(t\right)= \hat{\zeta}_j\left(0\right)e^{-i\beta_jt}. \label{Eq:OperatorsSolution}\ee 

By assigning eq.~\eqref{Eq:GammaInv} in eq.~\eqref{Eq:CommutationRelations}, we get that the canonical commutation relations of these operators obey
\be \left[\hat{\zeta}_j\left(t\right), \hat{\xi}_k\left(t\right)\right]=\delta_{jk} \label{Eq:CommutationRelations2}\ee
with all other commutators zero, and this result is subject to the normalization condition eq.~\eqref{Eq:Normalization} which ensures that the Floquet-Lyapunov transformation is canonical.
%The hermiticity of $\hat{\vec{u}}$ and $\hat{\vec{p}}$ immediately implies $\hat{\xi}_j^{\dag} = \hat{\zeta}_j$ (which can also be deduced directly from eq.~\eqref{Eq:Realness}).

The commutation relations of eq.~\eqref{Eq:CommutationRelations2} are easily recognizable as those of the creation and annihilation operators of a harmonic oscillator, since the hermiticity of $\hat{\vec{u}}$ and $\hat{\vec{p}}$ immediately implies $\hat{\xi}_j^{\dag} = \hat{\zeta}_j$ (which also follows directly from eq.~\eqref{Eq:Realness}). We may therefore define the time-independent eigenstates of $\hat{\zeta}\left(t\right)$, the coherent states, in the Heisenberg picture, by
\be \hat{\zeta}_j\left(t\right)\left|\zeta\right\rangle = \zeta_j\left(t\right)\left|\zeta\right\rangle \equiv \zeta_j\left(0\right) e^{-i\beta_j t}\left|\zeta\right\rangle, \label{Eq:CoherentState}\ee
and the normalization and completeness relations are
\be \left\langle { \zeta\left|\right.}\zeta'\right\rangle = e^{ \zeta^*\cdot \zeta'}, \qquad
\int d\mu_f \left| \zeta \right\rangle { \left\langle \zeta\right|}= \hat{1},\label{Eq:zetaNormalization}\ee
where $d\mu_f =\pi^{-f}e^{-\zeta^* \cdot \zeta}d^f\zeta$ and $d^f\zeta = \prod dx_j\prod dy_j$ if $\zeta$ is written in terms of the real variables $\zeta_j=x_j+iy_j$.

We next show that the Schr\"{o}dinger wavefunction of the coherent state vector of the system, is given in the coordinate representation by
\be \psi_{\zeta}\left(\vec{u},t\right)\equiv\left\langle{ \vec{u}\left|\right.}\zeta \right\rangle = \mathcal{N} \exp\left\{ i\tilde{\mathcal{F}}\left(\vec{u},\zeta,t\right)\right\}, \label{Eq:psi} \ee
where $\tilde{\mathcal{F}}\left(\vec{u},\zeta\left(t\right),t\right)$ has the same functional form of the classical generating function ${\mathcal{F}}\left(\vec{u},\zeta,t\right)$, only with the explicit time-dependence of $\zeta= \zeta\left(t\right)$, which we will omit below, except where necessary. Multiplying eq.~\eqref{Eq:CoherentState} by $\left\langle \vec{u}\right|$ we obtain the differential equation
\be \left(-i\sum_kV^t_{jk}{u}_k +\sum_kU^t_{jk}\partial_{u_k}\right) \left\langle{ \vec{u}\left|\right.} \zeta\right\rangle = {\zeta_j} \left\langle{ \vec{u}\left|\right.} \zeta \right\rangle.\label{Eq:MatrixElement}\ee
Since matrix elements are invariant under any unitary transformation, eq.~\eqref{Eq:MatrixElement} holds in the Schr\"{o}dinger picture as well. The solution of this equation is
\be \left\langle{ \vec{u}\left|\right.}\zeta \right\rangle = {\mathcal{N}}\exp\left\{\frac{1}{2}i \vec{u}^t U^{-t}V^{t}\vec{u} +\vec{u}^t U^{-t}\zeta + f\left(\zeta\right)\right\}.\ee

To determine the (time-dependent) normalization constant $\mathcal{N}$ and the function $f\left(\zeta\right)$, we first impose the normalization of eq.~\eqref{Eq:zetaNormalization},
\be e^{ \zeta^*\cdot \zeta'} = \left\langle { \zeta\left|\right.}\zeta'\right\rangle = \int_{\mathbb{R}^f} \left\langle \zeta\right.{\left|\vec{u} \right\rangle} \left\langle{ \vec{u}\left|\right.}\zeta' \right\rangle d^f\vec{u}.\label{Eq:NormalizationIntegral} \ee
%\be \delta\left( \vec{u}' - \vec{u}\right)=  \left\langle{ \vec{u}'\left|\right.}\vec{u}\right\rangle = \int_{\mathbb{C}^f} \left\langle{ \vec{u}'\left|\right.}\zeta\right) \left(\zeta\right.{\left|\vec{u} \right\rangle}  d{\mu}_f.\ee

To evaluate eq.~\eqref{Eq:NormalizationIntegral} we use the following result. Let $T$ be a symmetric $n\times n$ complex matrix with positive definite real part, and $b$ a complex vector. 
Then
\be \mathcal{J} = \int_{{\mathbb R}^n}{\exp\left\{-\frac{1}{2}x^tTx+b^tx\right\}d^nx}=\left(2\pi\right)^{n/2}\left(\det T\right)^{-1/2}\exp\left\{\frac{1}{2}b^tT^{-1}b\right\}, \ee
and the value of $\left(\det T\right)^{-1/2}$ is defined by analytic continuation, writing $T={\mathfrak Re}T+i\varepsilon{\mathfrak Im}T$ starting with the (positive definite) real part of $T$ and increasing $\varepsilon$ continously to $1$.

From eq.~\eqref{Eq:Identities} we get
\be U^{-t}V^t = U^{-\dag}V^{\dag} + iU^{-\dag}U^{-1},\label{Eq:IntegrationIdentities}\ee
so
\begin{widetext} \be e^{ \zeta^*\cdot \zeta'} = \left|\mathcal{N}\right|^2 \int_{\mathbb{R}^f} \exp\left\{ -\frac{1}{2}\vec{u}^tU^{-\dag}U^{-1}\vec{u} +\vec{u}^t U^{-t}\zeta' +\vec{u}^t U^{-\dag}\zeta^* + f\left(\zeta'\right) +f^*\left(\zeta\right) \right\}d^f\vec{u},\ee 
and we get (since $\left(U^{\dag}U\right)^{-1}$ is obviously positive definite),
\begin{multline} e^{ \zeta^*\cdot \zeta'} = \left|\mathcal{N}\right|^2 \left(2\pi\right)^{f/2} \det\left(UU^{\dag}\right)^{1/2}\times\\\times \exp\left\{ \frac{1}{2}\left(\zeta'^t U^{\dag}U^{-t}\zeta' +\zeta^{*t} U^{-*}U\zeta^* + \zeta^{*t}U^{-*}UU^{\dag}U^{-t}\zeta' + \zeta'^t\zeta^* \right)+ f\left(\zeta'\right) +f^*\left(\zeta\right)\right\}.\end{multline}
\end{widetext} 
By using the fact that $U^{\dag}U^{-t}$ is symmetric we find
\be \left(U^{-*}U\right)^*=U^{\dag}U^{-t}, \qquad U^{-*}UU^{\dag}U^{-t}=1_f, \label{Eq:SymmetricIdentities}\ee
so we may deduce
\be f\left(\zeta\right) =-\frac{1}{2}\zeta^t U^{\dag}U^{-t}\zeta, \quad \left|\mathcal{N}\right| = \left(2\pi\right)^{-f/4} \det\left(UU^{\dag}\right)^{-1/4}, \label{Eq:fNAbs}\ee
which proves eq.~\eqref{Eq:psi}, without fixing the phase of $\mathcal{N}$.

Finally, to determine the wavefunction completely, we require that $\psi_{\zeta}\left(\vec{u},t\right)$ be a solution of the Schr\"{o}dinger equation, which is expressed in coordindate $\vec{u}$ space using the Hamiltonian of eq.~\eqref{Eq:Hphi},
\be i\partial_t\psi_{\zeta}\left(\vec{u},t\right)= \hat{\mathcal H}\psi_{\zeta}\left(\vec{u},t\right) \label{Eq:psiSchrodinger}. \ee
Substituting eq.~\eqref{Eq:psi} in eq.~\eqref{Eq:psiSchrodinger} and inserting a resolution of the identity from eq.~\eqref{Eq:zetaNormalization} we have
\be i\left(\dot{\mathcal N}/{\mathcal N}+i\partial_t{\tilde{\mathcal F}}\right)\left\langle{ \vec{u}\left|\right.}\zeta \right\rangle = \int d\mu'_f \left\langle{ \vec{u}\left|\right.}\zeta' \right\rangle \left\langle \zeta'\right|\hat{\mathcal H}\left|\zeta\right\rangle. \label{Eq:psiCondition} \ee 
In App.~\ref{App:Hamiltonian} we show that the integral on the r.h.s of eq.~\eqref{Eq:psiCondition} is equal to
\be \left[ -\partial_t{\tilde{\mathcal F}}+\frac{1}{2}{\rm tr}\left\{B-iW\right\}\right]\left\langle{ \vec{u}\left|\right.}\zeta \right\rangle \ee
so that we get
\be \dot{\mathcal N}/{\mathcal N}= -\frac{1}{2}{\rm tr}\left\{iB+W\right\}.\ee

Writing ${\mathcal N}=\left|{\mathcal N}\right|\exp\left\{i\arg{\mathcal N}\right\}$, the above equation is equivalent to the two equations
\be \left|\dot{\mathcal N}\right|/\left|{\mathcal N}\right| = -\frac{1}{2}{\mathfrak Re}\left\{{\rm tr}W\right\}, \label{Eq:NdotAbs}\ee
and
\be \partial_t\arg{\mathcal N}=-\frac{1}{2}\left({\rm tr}B+{\mathfrak Im}\left\{{\rm tr}W\right\}\right).\label{Eq:NArgdot}\ee
Eq.~\eqref{Eq:NdotAbs} is in fact an identity which results from eq.~\eqref{Eq:fNAbs} as shown in App.~\ref{App:Hamiltonian}, where it is also shown that the solution of eq.~\eqref{Eq:NArgdot} is
\be \arg{\mathcal N}=-\frac{1}{2}\sum_j \beta_jt -\frac{1}{2}\arg\det U.\label{Eq:argN}\ee
%which gives \bm {\mathcal N}=\left(2\pi\right)^{-f/4} \det\left(UU^{\dag}\right)^{-1/4}\times \\ \times\exp\left\{-\frac{1}{2}i\sum_j \beta_jt -\frac{1}{2}i\arg\det U \right\}.\end{multline}

Let us write here again the complete expression for the coherent state vector $\psi_{\zeta}$,
\begin{widetext}
\begin{multline} \psi_{\zeta} = \left(2\pi\right)^{-f/4} \det\left(UU^{\dag}\right)^{-1/4}\times\\\times\exp\left\{-\frac{1}{2}i\sum_j \beta_jt -\frac{1}{2}i\arg\det U +\frac{1}{2}i \vec{u}^t U^{-t}V^{t}\vec{u} +\vec{u}^t U^{-t}\zeta -\frac{1}{2}\zeta^t U^{\dag}U^{-t}\zeta\right\},\label{Eq:psi_zeta_wide}\end{multline}
\end{widetext}
where $\zeta$ is the vector with components $\zeta_j\left(0\right)e^{-i\beta_jt}$.

In the case of a single harmonic oscillator of frequency $\beta$, eq.~\eqref{Eq:psi_zeta_wide} must reduce to the familiar wavefunction in configuration space of a coherent state, with complex label $\zeta_0\equiv\zeta\left(0\right)$. This can be seen by noting that for this case, the transformation matrices $U$ and $V$ with the normalization of eq.~\eqref{Eq:Normalization}, become the scalars $U=\frac{1}{\sqrt{2\beta}}$ and $V=i\sqrt{\frac{\beta}{2}}$, and then eq.~\eqref{Eq:psi_zeta_wide} becomes, with $m=1,\hbar=1$, the expected expression (e.g. eq.~(21.1.132) of \cite{Shankar1994})
\begin{equation*} \psi_{\zeta} = \left(2\pi\right)^{-1/4} \left(\frac{1}{2\beta}\right)^{-1/4}\exp\left\{-\frac{1}{2}i \beta t -\frac{1}{2}\beta {u}^2 +\sqrt{2\beta}u\zeta_0e^{-i\beta t} -\frac{1}{2}\zeta_0^2e^{-i2\beta t}\right\}.\end{equation*}

We will now find a complete orthonormal basis of solutions of the Schr\"{o}dinger equation. For that purpose we can use a generating function for multidimensional Hermite polynomials $H_{\vec{n}}^C$, defined by
\be G_C=\exp\left\{\vec{x}^tC\zeta-\frac{1}{2}\zeta^tC\zeta\right\}=\sum_{\vec{n}}\frac{\zeta_1^{n_1}}{n_1!}\cdots\frac{\zeta_f^{n_f}}{n_f!}H_{\vec{n}}^C\left(\vec{x}\right)\label{Eq:GeneratingFunction},\ee
where $C$ is a symmetric matrix, and the summation is over all $f$-tuples of nonnegative integers, $\vec{n}$. An explicit definition of $H_{\vec{n}}^C$ is given by \cite{Bateman}
\be H_{\vec{n}}^C\left(\vec{x}\right) = \left(-1\right)^{\sum n_j}e^{\vec{x}^tC\vec{x}/2}\frac{\partial^{n_1}}{\partial{x_1}^{n_1}}\cdots\frac{\partial^{n_f}}{\partial{x_f}^{n_f}}e^{-\vec{x}^tC\vec{x}/2}.\label{Eq:Hermite}\ee

 Then, setting in eq.~\eqref{Eq:GeneratingFunction} $\vec{x}=U^{-*}\vec{u}$ and
\be C=U^{\dag}U^{-t},\ee
 we can write $\psi_{\zeta}$ as
\be \psi_{\zeta}={\mathcal N} \exp\left\{\frac{1}{2}i \vec{u}^t U^{-t}V^{t}\vec{u}\right\}  \sum_{\vec{n}}e^{-i\sum_j n_j\beta_jt} \frac{\zeta_1\left(0\right)^{n_1}}{n_1!}\cdots\frac{\zeta_f\left(0\right)^{n_f}}{n_f!} H_{\vec{n}}^C\left(U^{-*}\vec{u}\right) \label{Eq:ZetaHermite}.\ee
Eq.~\eqref{Eq:ZetaHermite} can be interpreted as an expansion of $\psi_{\zeta}$ in terms of a complete orthonormal set of solutions of the Schr\"{o}dinger equation, $\psi_{\vec{n}}$, and in that case the coefficients of the expansion must be time-\textit{independent}. We may therefore write
\be \psi_{\vec{n}}={\mathcal N}_{\vec{n}} e^{-i\sum_j n_j\beta_jt}\exp\left\{\frac{1}{2}i \vec{u}^t U^{-t}V^{t}\vec{u}\right\}  H_{\vec{n}}^C\left(U^{-*}\vec{u}\right) \label{Eq:psi_n},\ee
where ${\mathcal N}_{\vec{n}}={c}_{\vec{n}}{\mathcal N}$ and ${c}_{\vec{n}}$ is time-independent, and we impose the normalization
\be \delta_{\vec{n},\vec{n}'} = \int \psi_{\vec{n}}^*\left(\vec{u},t\right)\psi_{\vec{n}'}\left(\vec{u},t\right)d^f\vec{u}. \label{Eq:deltann}\ee

In App.~\ref{App:Hamiltonian} we show that
\be {c}_{\vec{n}}= \left({n_1}!\cdots {n_f}!\right)^{-1/2} \label{Eq:N_n}\ee
so that we have
\begin{widetext}
\begin{multline} \psi_{\vec{n}}= \left(2\pi\right)^{-f/4}\left( \prod_j\frac{1}{\sqrt{n_j!}}\right)\det\left(UU^{\dag}\right)^{-1/4}\times\\\times\exp\left\{-i\sum_j\left( n_j+\frac{1}{2}\right) \beta_jt -\frac{1}{2}i\arg\det U +\frac{1}{2}i \vec{u}^t U^{-t}V^{t}\vec{u}\right\}H_{\vec{n}}^C\left(U^{-*}\vec{u}\right). \label{Eq:psi_n_wide}\end{multline}
\end{widetext}

Let us note how to get from eq.~\eqref{Eq:psi_n_wide} the familiar expression for the wavefunctions of the one-dimensional case, e.g. in the form of eq.~(36) of \cite{WinelandReview}. The latter is given in terms of the periodic function $\Phi\left(t\right)$ and the constant $\nu$ defined there, in eqs.~(22)-(23), with $\beta\equiv\beta_x$. Let us keep for now the nondimensional units (with the drive frequency being equal to 2), so $\nu=\sum \left(2n+\beta\right)C_{2n}$. By the normalization of eq.~\eqref{Eq:Normalization}, we see that $U\left(t\right)=\frac{1}{\sqrt{2\nu}}\Phi\left(t\right)$. The usual one-dimensional Hermite polynomials $H_n$ are obtained in eq.~\eqref{Eq:Hermite} by setting $C=2$. Since we have $C=\Phi^*/\Phi$, this requires the variable change
\[{z}=\sqrt{\frac{\Phi^*}{2\Phi}}{x},\qquad \frac{\partial}{\partial x}=\sqrt{\frac{\Phi^*}{2\Phi}}\frac{\partial}{\partial z}=2^{-1/2}e^{-i\arg\Phi}\frac{\partial}{\partial z},\]
which gives that
\[ H_{\vec{n}}^C\left(U^{-*}\vec{u}\right) = \frac{e^{-in\arg\Phi}}{\sqrt{2^n}}H_n\left(\sqrt{\frac{\nu}{\left|\Phi\right|^2} } u\right).\]
In addition we have 
\[\left(2\pi\right)^{-1/4}\left(\det UU^{\dag}\right)^{-1/4}e^{-\frac{1}{2}i\arg\det U}=\frac{\left(\nu/\pi\right)^{1/4}}{\Phi^{1/2}},\]
and $\frac{1}{2}iU^{-t}V^t=i\frac{\dot{\Phi}}{\Phi}-\beta$. Eq.~\eqref{Eq:eomuHomogenousVec} is expressed in terms of rescaled time, and we now return to the time variable before the Mathieu scaling $t\to\Omega t/2$ (with $\Omega$ being the physical drive frequency), and therefore put $\nu\to \nu\Omega/2$, and put back $\hbar$ (we still have $m=1$), to get the wavefunction \footnote{We note that eq.~(36) of \cite{WinelandReview} contains a misprint in the sign of the coefficient of $x'^2$, and where $\nu$ appears instead of $\beta_x$ inside the two $\exp$ factors, as can be verified from eq.~(34) of \cite{WinelandReview}, or from \cite{Glauber1992}.}
\begin{widetext}
\be \psi_{{n}}= \frac{e^{-in\arg\Phi}}{\sqrt{2^n n!}\Phi^{1/2}} \left(\frac{\nu}{\pi\hbar}\right)^{1/4}\exp\left\{-i\left( n+\frac{1}{2}\right) \beta_x\frac{\Omega}{2} t +\frac{1}{2\hbar}\left(i\frac{\dot{\Phi}}{\Phi}-\beta_x\frac{\Omega}{2}\right) u^2\right\}H_{{n}}\left(\sqrt{\frac{\nu}{\hbar\left|\Phi\right|^2} } u\right). \label{Eq:psi_n_dimensional}\ee
\end{widetext}

In a multidimensional problem, two distinct situations may arise. If $U$ is diagonal, which means that $C$ is diagonal as well, the generating function of eq.~\eqref{Eq:GeneratingFunction} obviously factorizes into a product of exponents, and the wavefunction will be a product of one-dimensional wavefunctions, each depending exclusively on one variable, as obtained above. As mentioned in Sec.~\ref{Sec:Floquet}, $U$ can be made be diagonal if there exists a constant matrix which diagonalizes the equations of motion. Then $U$ will be diagonal in some normal modes which are time-independent linear combinations of the original coordinates. If such a diagonalization does not exist, the wavefunctions will depend on (time-dependent) complex linear combinations of the coordinates, through the multidimensional Hermite polynomials.

\subsection{The Inhomogenous Equations}\label{Sec:Inhom}

In this subsection we describe briefly how to obtain the wavefunctions of the quantum system which corresponds to eq.~\eqref{Eq:eomuHomogenousVec} with a driven r.h.s in the form
\be\ddot{\vec{u}} +\left[A -2Q \cos 2t\right]\vec{u} = \vec{G} +2\vec{F}\cos2t,\label{Eq:eomuInHomogenous2}\ee 
where $\vec{G}$ and $\vec{F}$ are $f$-component constant vectors. We rewrite eq.~\eqref{Eq:eomuInHomogenous2} in Floquet form (eq.~\eqref{Eq:Pioft}) using
\[\lambda =\left(\begin{array}{c} {0} \\ {\vec{G} +2\vec{F}\cos2t} \end{array}\right) \]
as
\be\dot{\phi } -\Pi \left(t\right) \phi =\lambda \left(t\right), \ee 
which we transform using the Floquet-Lyapunov transformation eq.~\eqref{Eq:FLTrans} and its inverse, to get the e.o.m in the form
%\be\dot{\chi }_{\nu } -i\beta _{\nu } \chi _{\nu } = \left(\Gamma \left(t\right)^{-1} \right)_{\nu\tau } \lambda _{\tau }, \label{Eq:eomchi}\ee 
%or
\be\dot{\chi }-{\rm B} \chi =\Gamma \left(t \right)^{-1} \lambda. \label{Eq:eomchiInHomogenous}\ee 

Each of the equations, eq.~\eqref{Eq:eomchiInHomogenous} and eq.~\eqref{Eq:eomuInHomogenous2}, has a unique $\pi$-periodic solution (see App.~\ref{App:Inversions}), and these solutions are related by the Floquet-Lyapunov transformation. Eq.~\eqref{Eq:eomchiInHomogenous} can be solved immediately term-by-term, however in App.~\ref{App:Inversions} we use infinite continued matrix inversions to find directly the periodic solution of eq.~\eqref{Eq:eomuInHomogenous2}, in the form 
\be \vec{u}_{\pi}=\sum _{n\in {\mathbb Z}}\vec{B}_{2n} e^{i2nt}\label{Eq:uInhomSol}.\ee

To get the wavefunctions, we use the method of \cite{Husimi1953}. The Schr\"{o}dinger equation in coordindate $\vec{u}$ space is now
\be i\partial_t{\psi}=-\nabla_{\vec{u}}^2\psi+\left[\frac{1}{2}\vec{u}^t\left(A-2Q\cos 2t\right)\vec{u}-\left(\vec{G} +2\vec{F}\cos2t\right)\cdot\vec{u}\right]\psi.\label{Eq:SchrodingerDriven}\ee
Performing the (unitary) coordinate change $\vec{x}=\vec{u}-\vec{u}_{\pi}$ we have $\partial_t\to-\dot{\vec{u}}_{\pi}+\partial_t$, and by introducing the additional unitary transofmration,
\be \psi = \exp\left\{i\dot{\vec{u}}_{\pi}\cdot \vec{x}+i\alpha_{\pi}\left(t\right)\right\}\varphi,\ee
the Schr\"{o}dinger equation for $\varphi$ becomes that of the nondriven problem, whose solutions are given in eq.~\eqref{Eq:psi_zeta_wide} and eq.~\eqref{Eq:psi_n_wide}. The phase $\alpha_{\pi}\left(t\right)$ is in fact the classical action of the $\pi$-periodic solution,
\be \alpha_{\pi}\left(t\right) = \int_0^t{\frac{1}{2}\left(\dot{\vec{u}}_{\pi}\right)^2} -\int_0^t{\left[\frac{1}{2}\vec{u}_{\pi}^t\left(A-2Q\cos 2t\right)\vec{u}_{\pi}-\left(\vec{G} +2\vec{F}\cos2t\right)\cdot\vec{u}_{\pi}\right]},\ee
which may be written more compactly using eq.~\eqref{Eq:eomuInHomogenous2},
\be \alpha_{\pi}\left(t\right) = \int_0^t{\frac{1}{2}\left[\left(\dot{\vec{u}}_{\pi}\right)^2+{\vec{u}}_{\pi}^t\cdot\left(\ddot{\vec{u}}_{\pi}+\vec{G}+2\vec{F}\cos2t\right)\right]}.\label{Eq:phipi}\ee

A closed algebraic expression for $\alpha_{\pi}\left(t\right)$ can in fact be obtained by expanding the integrand of eq.~\eqref{Eq:phipi} using eq.~\eqref{Eq:uInhomSol} into a sum of exponentials, for which the integration is immediate.

Thus, with $\varphi\left(\vec{x},t\right)$ a solution of the nondriven Schr\"{o}dinger equation, the solutions of eq.~\eqref{Eq:SchrodingerDriven} will be
\be \psi=\exp\left\{i\dot{\vec{u}}_{\pi}\cdot \left(\vec{u}-\vec{u}_{\pi}\right)+i\alpha_{\pi}\left(t\right)\right\}\varphi\left(\vec{u}-\vec{u}_{\pi},t\right).\ee

%\be\chi _{\nu } =A_{\nu } \left(t \right)e^{i\beta _{\nu } t} +\left[\Gamma^{-1} \left(t\right) \left(\begin{array}{c} {\sum \vec{B}_{2n} e^{i2nt}} \\ {\sum i2n \vec{B}_{2n} e^{i2nt}} \end{array}\right) \right]_{\nu } \ee

\section{Concluding Comments}\label{Sec:Conclusion}

Eq.~\eqref{Eq:eomuHomogenousVec} describes a coupled system of Mathieu equations. This equation can be considered as consisting of the first two terms in the expansion in a Fourier series of a more general system of coupled Hill equations \cite{McLachlan},
\be\ddot{\vec{u}} +\left[A -2\sum_{n=1} Q_{2n}\cos 2nt\right]\vec{u} = 0,\label{Eq:eomHill} \ee 
where the sum may be infinite in principle, and here we take the equation to be time-reversal invariant. The technique of expansion of Sec.~\ref{Sec:MathieuSolution} in matrix inversions can be applied to solve eq.~\eqref{Eq:eomHill} with only the algebraic overhead growing \cite{Risken}. The Floquet-Lyapunov transformation and the entire quantum treatment remain identical. In App.~\ref{App:DoubleCos} we solve a Hill system with two harmonics.

Beside the above generalization, we have considered in App.~\ref{App:Inversions} an inhomogenous system (of a specific form) and its quantum counterpart is considered in Sec.~\ref{Sec:Inhom}. Other types of driving can be handled similarly, and simple known transformations \cite{Yakubovich} can be used to handle more general linear system similar to eq.~\eqref{Eq:eomHill}, such as systems with first-order derivatives (e.g. linear damping, gyroscopic forces or magnetic fields). 

Finally, we note that a linearization of a general nonlinear multidimensional system with periodic coefficients will lead to similar equations. After obtaining the Floquet-Lyapunov transformation, the system of decoupled time-independent oscillators can be canonically transformed to action-angle coordinates, for example. The obtained modes can be used as the zeroth-order approximation in a perturbative treatment of the nonlinearity \cite{PandiyanSinha,ButcherSinha}.

An application of the methods described here to the analysis of the classical linearized modes of an ion crystal in a Paul trap can be found in \cite{rfions}, where the accuracy of the solution is demonstrated by comparing to exact numerical simulations of the nonlinear problem. Various other linear and nonlinear parametrically driven physical systems can be accurately described and analyzed, either in the classical regime or as they are cooled close to the quantum ground-state of motion. Coupled arrays of nanoelectromechanical oscillators \cite{LifshitzParametricOscillators,LifshitzNonlinearReview,LifshitzNonlinearReview2} are one example. Parametric driving has also been recently applied to Bose-Einstein condensates \cite{BECFaradayWaves,BECFaradayWavesTheory,BinaryBEC}, and in particular the perturbations (i.e. phonons) of two-component condensates obey coupled Mathieu equations \cite{BECCoupledMathieu}.

\section*{Acknowledgments}

BR acknowledges the support of the Israel Science Foundation. BR and AR acknowledges the support of the German-Israeli Foundation and the EU STReP project PICC. MD acknowledges financial support from the Carlsberg Foundation and the EU via the FP7 projects `Physics of Ion Coulomb Crystals' (PICC) and `Circuit and Cavity Quantum Electrodynamics' (CCQED). HL wishes to thank R. Geffen.

\begin{appendix}

\section{Further Comments on The Infinite Continued Matrix Inversions}\label{App:Inversions}

Adding to eq.~\eqref{Eq:eomuHomogenousVec} an inhomogeneous r.h.s we put it in the form
\be\ddot{\vec{u}} +\left[A -2Q \cos 2t\right]\vec{u} = \vec{G} +2\vec{F}\cos2t,\label{Eq:eomuInHomogenous}\ee 
where $\vec{G}$ and $\vec{F}$ are $f$-component constant vectors. This equation will have a unique $\pi$-periodic solution under some conditions \cite{Yakubovich}, a sufficient condition being that the homogenous equation does not have any $\pi$-periodic solution (except the trivial one). This condition is of course fulfilled when the homogenous system has purely oscillatory modes.
We first find a particular solution of eq.~\eqref{Eq:eomuInHomogenous}, using the method of Sec.~\ref{Sec:MathieuSolution}. We assign $u_{\pi}=\sum _{n\in {\mathbb Z}}\vec{B}_{2n} e^{i2nt} $ in the e.o.m and get
\be \left(A-4n^{2} \right)\vec{B}_{2n} -Q\left(\vec{B}_{2n-2} +\vec{B}_{2n+2} \right)=\vec{G}\delta _{n,0} +\vec{F}\left(\delta _{n,1} +\delta _{n,-1} \right).\ee
We write, defining $R_{2n} =A-(2n)^{2}$ and using $\vec{B}_{2n} =\vec{B}_{-2n}$,
\be A\vec{B}_{0} -2Q\vec{B}_{2} =\vec{G}\label{Eq:B0}\ee 
\be R_2\vec{B}_{2} -Q\left(\vec{B}_{0} +\vec{B}_{4} \right)=\vec{F}\label{Eq:B2}\ee 
\be R_{2n}\vec{B}_{2n} -Q\left(\vec{B}_{2n-2} +\vec{B}_{2n+2} \right)=0,\quad n\ge 2.\label{Eq:B2n}\ee 
Eq.~\eqref{Eq:B2n} immediately gives a recursion relation in the form of eq.~\eqref{Eq:Recursion1} (only $R_{2n}$ here is defined differently), which allows us to obtain the expression in infinite inversions
\be \vec{B}_{4}=T_2Q\vec{B}_{2},\label{Eq:B4}\ee
where
\be T_2= \left[R_4-Q\left[R_6-Q\left[R_8-...\right]^{-1}Q\right]^{-1}Q\right]^{-1}.\ee
Substituting eq.~\eqref{Eq:B4} in eqs.~\eqref{Eq:B0}-\eqref{Eq:B2} we get the linear system
\be \left(\begin{array}{cc} {A} & {-2Q } \\ {-Q} & {R_2-QT_2Q} \end{array}\right)\left(\begin{array}{c} {\vec{B}_0} \\ {{\vec{B}_2}} \end{array}\right) =\left(\begin{array}{c} {\vec{G}} \\ {{\vec{F}}} \end{array}\right), \ee
which is readily solved, and the rest of the coefficients follow immediately.

%\be \vec{B}_{2}=T_2\left(\vec{F}+QA^{-1}\vec{G}\right)\label{Eq:B2Solution}\ee
%where
%\bm T_2=\\ \left[R_2-2QA^{-1}Q-Q\left[Q^{-1}R_4-\left[Q^{-1}R_6-...\right]^{-1}\right]^{-1}\right]^{-1}.\end{multline}
%Then we have
%\be \vec{B}_{0} =A^{-1} \left(\vec{G}+2Q\vec{B}_{2} \right)\label{Eq:B0},\ee 
%which we substitute in eq.~\eqref{Eq:B2} to obtain
%\be \left(R_2-2QA^{-1}Q\right)\vec{B}_{2} -Q\vec{B}_{4}=\vec{F}+QA^{-1}\vec{G}.\label{Eq:B2b}\ee

We conclude by commenting on the computational aspects of the above method. The complexity of matrix multiplications and inversion, and computing a determinant are all equal, ${\rm O}\left(f^3\right)$ or a bit better with improved algorithms. Using the method of matrix inversions, finding all zeros of the determinant of eq.~\eqref{Eq:Y2beta} can be done with complexity independent of $f$, albeit with a large constant prefactor, or at ${\rm O}\left(f\log f\right)$ operations. A brute-force computation of the monodromy matrix by repeated integrations of the e.o.m would have complexity ${\rm O}\left(f^4\right)$ because $f$ passes are required in order to obtain $f$ linearly independent solutions. Approximations have been developed which yield a fundamental matrix solution in ${\rm O}\left(f^3\right)$. From the matrizant the characteristic exponents can be obtained. In \cite{Takahashi}, and similarly in \cite{Hansen1985}, following earlier works, a solution to recursion relations (similar to, e.g., eq.~\eqref{Eq:Recursion1}) is achieved by truncating a linear eigenvalue problem. The convergence of the expansion using continued inversions is expected to be much better (this is certainly true for a single degree of freedom). In \cite{Sinha}, an expansion using Chebyshev polynomials is used to approximate a fundamental matrix solution.

\section{Solution of the Double-Cosine System}\label{App:DoubleCos}

As discussed in Sec.~\ref{Sec:Conclusion}, the expansion in continued inversions can be extended to treat coupled systems of Hill equations, and here we consider the double-cosine system
\be\ddot{\vec{u}} +\left[A -2 Q_{2}\cos 2t-2 Q_{4}\cos 4t\right]\vec{u} = 0.\label{Eq:eomDoubleCos} \ee 
By substitution of the solution ansatz, eq.~\eqref{Eq:uAnsatz}, we get the identity
\be R_{2n}C_{2n}-Q_2\left(\vec{C}_{2n-2}+\vec{C}_{2n+2}\right)-Q_4\left(\vec{C}_{2n-4}+\vec{C}_{2n+4}\right)=0, \label{Eq:RecursionIdentity}\ee
which gives the two recursion relations,
\be Q_4\vec{C}_{2n-4} =-Q_2\vec{C}_{2n-2}+R_{2n} \vec{C}_{2n} - Q_2\vec{C}_{2n+2}- Q_4\vec{C}_{2n+4}, \label{Eq:RecursionD1}\ee 
and
\be Q_2\vec{C}_{2n+2} =-Q_4\vec{C}_{2n+4}+R_{2n} \vec{C}_{2n} - Q_2\vec{C}_{2n-2}- Q_4\vec{C}_{2n-4}. \label{Eq:RecursionD2}\ee 
We do not obtain the expansion to the general order, but suffice with assigning $n=1,2,3$ in eq.~\eqref{Eq:RecursionD1} and $n=0,-1,-2$ in eq.~\eqref{Eq:RecursionD2} to get two expression in a form which is a generalization of eqs.~\eqref{Eq:MatrixInversions1}-\eqref{Eq:MatrixInversions2},
\be \vec{C}_{2} =T_{2,\beta }\tilde{Q}_{2,\beta} \vec{C}_{0}, \qquad\tilde{Q}_{-2,\beta}\vec{C}_{2} = \tilde{T}_{0,\beta }\vec{C}_{0} \label{Eq:C0C2D}\ee
with
\begin{widetext}
\be T_{2,\beta } =\left[R_2-Q_2\tilde{R}_4\tilde{R}_6-Q_4 R_6^{-1} Q_4-Q_4 R_6^{-1}Q_2\tilde{R}_4\tilde{R}_6-Q_4\tilde{R}_{-2}Q_{4}\right]^{-1},\ee
\be \tilde{Q}_{2,\beta}=\left[Q_2+Q_2\tilde{R}_4 Q_4+Q_4 R_6^{-1}Q_2\tilde{R}_4 Q_4 +Q_4\tilde{R}_{-2}\tilde{R}_{-4}\right],\ee 
and
\be \tilde{T}_{0,\beta } =\left[R_0-Q_2\tilde{R}_{-2}\tilde{R}_{-4} - Q_4 R_{-4}^{-1}Q_4 -Q_4{R}_{-2}^{-1}Q_2\tilde{R}_{-2}\tilde{R}_{-4}-Q_4\tilde{R}_{4}Q_4\right],\ee 
\be \tilde{Q}_{-2,\beta} =\left[Q_2+Q_2\tilde{R}_{-2}Q_4+Q_4R_{-4}^{-1}Q_2\tilde{R}_{-2}Q_4+Q_4\tilde{R}_4\tilde{R}_6\right],\ee
and we have defined the matrices
\be \tilde{R}_4=\left[R_4-Q_2 R_6^{-1} Q_2\right]^{-1},\qquad\tilde{R}_6=\left[Q_2+Q_2 R_6^{-1} Q_4\right],\ee
\be \tilde{R}_{-2}=\left[R_{-2}-Q_2 R_{-4}^{-1} Q_2\right]^{-1},\qquad\tilde{R}_{-4}=\left[Q_2+Q_2 R_{-4}^{-1} Q_4\right].\ee
\end{widetext}
The characteristic exponents are obtained as zeros of the determinant of
\be Y_{2,\beta } \equiv \tilde{T}_{0,\beta } - \tilde{Q}_{-2,\beta } T_{2,\beta } \tilde{Q}_{2,\beta }, \ee
which is the generalization of eq.~\eqref{Eq:Y2beta}, and in fact reduces to it for $Q_4=0$. The mode vector $\vec{C}_{2}$ follows from eq.~\eqref{Eq:C0C2D}, and the other three vectors in this finite expansion can be obtained from
\be \vec{C}_{4} =\tilde{R}_{4}\left(Q_4\vec{C}_{0} + \tilde{R}_{6} \vec{C}_{2}\right), \ee
\be \vec{C}_{-2} =\tilde{R}_{-2}\left(\tilde{R}_{-4} \vec{C}_{0} + Q_4\vec{C}_{2}\right), \ee
\be \vec{C}_{-4} =\left[{R}_{-4}\right]^{-1}\left(Q_2 \vec{C}_{-2} + Q_4\vec{C}_{0}\right). \ee

\section{Proof of Several Results from Sec.~\ref{Sec:Quantization}}\label{App:Hamiltonian}

In this appendix we prove some results used in Sec.~\ref{Sec:Quantization}. Let us first obtain an explicit expression for the inverse of the Floquet-Lyapunov transformation. The matrizant $\Phi\left(t\right)$ of eq.~\eqref{Eq:eomCanonical} satisfies the identity
\be\Phi\left(t\right)^{\dag}{\rm J}\Phi\left(t\right)={\rm J}, \label{Eq:MatrizantInvariance}\ee 
as can be shown by differentiating and using eq.~\eqref{Eq:eomCanonical} (the matrizant of eq.~\eqref{Eq:eomHamiltonian} satisfies a similar identity with ${\rm J}$ replaced by ${\rm K}$). Multiplying eq.~\eqref{Eq:MatrizantInvariance} on the left with ${\rm J}$ and rearranging a little we find
\[ \Phi\left(t\right)^{-1} = -{\rm J}\Phi\left(t\right)^\dag{\rm J}.\]
From eq.~\eqref{Eq:Phit} we have that
\[\Gamma \left(t\right)^{-1} =e^{{\rm B} t} \Gamma ^{-1} \left(0\right) \Phi ^{-1} \left(t\right).\] 
Using these two expressions and substituting eq.~\eqref{Eq:Phit} we get
\be \Gamma \left(t\right)^{-1} = -e^{{\rm B}t}\Gamma^{-1}\left(0\right){\rm J}\Gamma^{-\dag}\left(0\right) e^{-{\rm B}t}\Gamma ^{\dag} \left(t\right){\rm J}, \label{Eq:AppGammaInv}\ee
 where $\Gamma ^{-\dag} \equiv \left[\Gamma ^{-1} \right]^{\dag}$ as was defined already above. By substituting $t \to t + T$ in the above equation and using the periodicity of $\Gamma$ and the fact the ${\rm B}$ is diagonal, we find that
\be \left[\Gamma ^{-1} \left(0\right){\rm J} \Gamma^{-\dag} \left(0\right),{\rm B}\right] = 0 \label{Eq:Bcommutator}\ee 
where $\left[ ,\right]$ is the matrix commutator.
% !! seems to be some kind of inverse of 1/2 B which might be 1/Omega B ??
We can find the explicit form of $\Gamma ^{-1} \left(0\right)$ using the fact that $U\left(0\right)$ is real and $V\left(0\right)$ is purely imaginary,
\[ \Gamma ^{-1} \left(0\right) = \frac{1}{2} \left(\begin{array}{cc} {U^{-1}\left(0\right) } & {V^{-1}\left(0\right) } \\ {U^{-1}\left(0\right) } & {-V^{-1}\left(0\right) }\end{array}\right), \]
so,
\be \Gamma ^{-1} \left(0\right){\rm J}\Gamma ^{-\dag} \left(0\right) = \left(\begin{array}{cc} {M +M^t } & {M^t-M} \\ {M-M^t} & {-M-M^t}\end{array}\right), \ee
with
\be M = \frac{1}{4}\left[ U^{-1}\left(0\right) V^{-t}\left(0\right)\right] \label{Eq:M}. \ee
Expanding eq.~\eqref{Eq:Bcommutator} in block form, using eq.~\eqref{Eq:Bblock}, we have, writing $\left\{,\right\}_+$ for the matrix anti-commutator,
\be \left[M+M^t,B\right] = 0, \qquad \left\{M-M^t,B\right\}_+ = 0, \ee
which together imply
\be \left[M,B\right] = 0. \label{Eq:Mcommutator} \ee

Eq.~\eqref{Eq:Mcommutator} means that the invariant subspaces of $M$ and $B$ are identical, and since $B$ is diagonal, we can conclude that $M$ must be diagonal too (or, if $B$ has degenerate eigenfrequencies, $M$ can be made diagonal by an appropriate choice of eigenvectors). We can make $M$ a scalar matrix, by demanding a proper normalization of $U$ and $V$. As noted at the end of Sec.~\ref{Sec:FLTrans}, the choice in eq.~\eqref{Eq:Gamma} is unique up to a matrix that commutes with ${\rm B}$, which amounts to the arbitrariness of the normalization of each of the columns of $U$, i.e. the fact that each of the vectors $\vec{C}_{0,\beta_j}$ was determined only up to a constant. This allows to to rescale and $U$ and $V$ as in eq.~\eqref{Eq:ScalingMatrix} and obtain eq.~\eqref{Eq:GammaInv} and eq.~\eqref{Eq:Identities}.

Let us find the transformed Hamiltonian of eq.~\eqref{Eq:Hamiltonian2},
\be {\mathcal H}'\left(\zeta,\xi\right)={\mathcal H}\left(\vec{u}\left(\zeta,\xi\right),\vec{p}\left(\zeta,\xi\right)\right) + \dot{\mathcal F}\left(\vec{u}\left(\zeta,\xi\right),\zeta\right).\label{Eq:TransformedH}\ee
By use of eq.~\eqref{Eq:Identities} it follows that
\be\Gamma^t{\rm J}\Gamma=\left(\begin{array}{cc} {0} & {i} \\ {-i} & {0}\end{array}\right),\label{Eq:GtJG}\ee
and by derivating this equation we get
\be\dot{\Gamma}^t{\rm J}\Gamma+\Gamma^t{\rm J}\dot{\Gamma}=0.\label{Eq:dGJG_GtJdG}\ee
Substituting $\phi=\Gamma\chi$ in eq.~\eqref{Eq:Hphi} we have
\be {\mathcal H} = \frac{1}{2} \chi^t\Gamma^t{\rm J}\Pi\Gamma\chi= \frac{1}{2} \chi^t\Gamma^t{\rm J}\left(\Gamma{\rm B}+\dot{\Gamma}\right)\chi= \frac{1}{2}\chi^t {\rm \tilde{H}} \chi - \frac{1}{2} \chi^t\dot{\Gamma}^t{\rm J}\Gamma\chi.\label{Eq:iH}\ee
where eqs.~\eqref{Eq:GtJG}-\eqref{Eq:dGJG_GtJdG} have been used to get the second line, and $\tilde{\rm H}$ is given by eq.~\eqref{Eq:tildeH}. We write explicitly
\be \dot{\Gamma}^t{\rm J}\Gamma = \left(\begin{array}{cc} {P} & {Q} \\ {R} & {T} \end{array}\right)\equiv \left(\begin{array}{cc} {\dot{V}^{t}U-\dot{U}^{t}V} & {\dot{V}^{t}U^*-\dot{U}^{t}V^*} \\ {\dot{V}^{\dag}U-\dot{U}^{\dag}V} & {\dot{V}^{\dag}U^*-\dot{U}^{\dag}V^*}\end{array}\right).\label{Eq:dGtJG}\ee
 By applying $\partial_t$ to eq.~\eqref{Eq:F} and substituting $\vec{u}=U\xi+U^*\zeta$, we may write after some rearranging
\be \dot{{\mathcal F}}=\frac{1}{2}\chi^t\Lambda\chi\label{Eq:dF}\ee
where
\be \Lambda=\left(\begin{array}{cc} {U^tLU} & {U^{t}LU^*-2iU^{t}\dot{U}^{-t}} \\ {U^{\dag}LU} & {U^{\dag}LU^* + i\dot{U}^{\dag}U^{-t}-iU^{\dag}\dot{U}^{-t}}\end{array}\right)\ee
and
\be L = \dot{U}^{-t}V^{t}+{U}^{-t}\dot{V}^{t}, \ee
and in these expressions, it is understood that $\dot{U}^{-t}\equiv\partial_t\left({U}^{-t}\right)$, i.e. the time-derivative is applied after matrix inversion. From eq.~\eqref{Eq:CanonicalCondition} we obtain the identity
\be \dot{V}^{\dag}-\dot{U}^{\dag}{U}^{-t}V^{t}-U^{\dag}L=-i\dot{U}^{-1},\ee
and by derivating $UU^{-1}=1_f$ we get in addition
\be {U}^{t}\dot{U}^{-t}=-\dot{U}^{t}{U}^{-t}. \ee
Using these identities and eq.~\eqref{Eq:Identities}, $\Lambda$ is simplified to
\be \Lambda=\frac{1}{2}\dot{\Gamma}^t{\rm J}\Gamma +\frac{1}{2} \left(\begin{array}{cc} {0} & {-iU^t\dot{U}^{-t}} \\ {i\dot{U}^{-1}U} & {i\dot{U}^{-1}U^*-iU^{\dag}\dot{U}^{-t}}\end{array}\right) \label{Eq:Lambda}.\ee
Since $\left(\dot{U}^{-1}U^*\right)^t=U^\dag\dot{U}^{-t}$, the lower-right block of the second term above is antisymmetric and therefore the coefficient of $\zeta^t\zeta$ will equal $0$. By using eqs.~\eqref{Eq:Lambda}, \eqref{Eq:dF} and \eqref{Eq:iH} in eq.~\eqref{Eq:TransformedH} we get finally
\be {\mathcal H}' = \frac{1}{2}\chi^t {\rm \tilde{H}} \chi + {\mathcal Y}\left(t\right)\label{Eq:TransformedHfinal}\ee
with 
\be {\mathcal Y} = -\frac{1}{2}i\left(\zeta^t W \xi - \xi^t W^t\zeta\right), \qquad W = {U}^{-1}\partial_tU.\label{Eq:Y}\ee
Classically, since $\zeta$ and $\xi$ commute, ${\mathcal Y}=0$.
 
We now evaluate the integral on the r.h.s of eq.~\eqref{Eq:psiCondition},
\be {\mathcal J} = \int d\mu'_f \left\langle{ \vec{u}\left|\right.}\zeta' \right\rangle \left\langle \zeta'\right|\hat{\mathcal H}\left|\zeta\right\rangle. \ee 
Using eq.~\eqref{Eq:iH} we can write $\hat{\mathcal H}$ as a function of $\hat{\xi}$ and $\hat{\zeta}$ and normal-order it (i.e. put all $\hat{\xi}$'s on the left and $\hat{\zeta}$'s on the right) to get 
\be \hat{\mathcal H} = \hat{\xi}^tB\hat{\zeta} - :\frac{1}{2}\hat{\chi}^t \dot{\Gamma}^t{\rm J}\Gamma \hat{\chi}: -\frac{1}{2}{\rm tr}\left\{R-B\right\}\label{Eq:NormalOrderedH}\ee
where $R$ appears in the lower-left block of eq.~\eqref{Eq:dGtJG}.
%and $W$ is defined in eq.~\eqref{Eq:Y}.

For any operator $\hat{\mathcal O}\left(\hat{\xi},\hat{\zeta}\right)$ we have
\[ \left\langle \zeta'\right|:\hat{\mathcal O}\left(\hat{\xi},\hat{\zeta}\right):\left|\zeta\right\rangle = \left\langle \zeta'\right|{\mathcal O}\left(\zeta'^*,{\zeta}\right) \left|\zeta\right\rangle = {\mathcal O}\left(\zeta'^*,{\zeta}\right) e^{\zeta'^*\cdot\zeta},\]
and we use the following result
\be \int f\left(\zeta'\right) \left[\zeta_j'^*\right]^n e^{\zeta'^*\cdot\zeta}d\mu'_f = \frac{\partial^n f\left(\zeta\right)}{\partial{\zeta_j}^n}.\ee 
The first term in eq.~\eqref{Eq:NormalOrderedH} gives
\be \int d\mu'_f \left\langle{ \vec{u}\left|\right.}\zeta' \right\rangle \left\langle \zeta'\right| \hat{\xi}^tB\hat{\zeta}\left|\zeta\right\rangle = \sum_{jk}\left( \partial_{\zeta_j}e^{i\tilde{\mathcal F}}\right)B_{jk}\zeta_k=\left[ \zeta^t U^{\dag}U^{-t}B{\zeta} -\vec{u}^tU^{-t}B{\zeta}\right]\left\langle{ \vec{u}\left|\right.}\zeta \right\rangle. \ee
We next evaluate the upper-left block of the second term in eq.~\eqref{Eq:NormalOrderedH},
\be -\frac{1}{2}\int d\mu'_f \left\langle{ \vec{u}\left|\right.}\zeta' \right\rangle \left\langle \zeta'\right| \hat{\xi}^tP\hat{\xi}\left|\zeta\right\rangle = -\frac{1}{2}\sum_{jk}P_{jk}\left[\left( i\partial_{\zeta_j}\tilde{\mathcal F}\right)\left( i\partial_{\zeta_k}\tilde{\mathcal F}\right)+i\partial_{\zeta_j}\partial_{\zeta_k}\tilde{\mathcal F}\right]e^{i\tilde{\mathcal F}} \label{Eq:IntP}. \ee 
By the definition of the trace and the identity in eq.~\eqref{Eq:CanonicalCondition} we have for the second term above,
\begin{multline} -\frac{1}{2}\sum_{jk}P_{jk}i\partial_{\zeta_j}\partial_{\zeta_k}\tilde{\mathcal F}=\frac{1}{2}\sum_{jk}\left(\dot{V}^tU-\dot{U}^tV\right)_{jk}\left(U^{\dag}U^{-t}\right)_{jk} =\\=\frac{1}{2}{\rm tr}\left\{U^{\dag}U^{-t}\left(\dot{V}^tU-\dot{U}^tV\right)^t\right\}=\frac{1}{2}{\rm tr}\left\{U^{\dag}\dot{V}-V^{\dag}\dot{U}-iU^{-1}\dot{U}\right\}=\frac{1}{2}{\rm tr}\left\{R-iW\right\}\end{multline}

By using the formal identity resulting from eq.~\eqref{Eq:GeneratingConditions},
\be \partial_{\zeta_j}e^{i\tilde{\mathcal F}}=\left(i\partial_{\zeta_j}\tilde{\mathcal F}\right)e^{i\tilde{\mathcal F}}=\xi_je^{i\tilde{\mathcal F}},\ee
we can immediately see that the first term in eq.~\eqref{Eq:IntP} and the remaining terms eq.~\eqref{Eq:NormalOrderedH} integrate to give simply $-\dot{\mathcal F}$, and together with $\dot{\zeta}=-iB\zeta$ we can collect all the terms to get that 
\be {\mathcal J} = \left[ -\partial_t{\tilde{\mathcal F}} +\frac{1}{2}{\rm tr}\left\{B-iW\right\}\right] \left\langle{ \vec{u}\left|\right.}\zeta \right\rangle. \ee

%\bm \left[ {\mathcal H}'\left(\zeta'^*,{\zeta}\right) -\dot{\mathcal F}\left(\zeta'^*,{\zeta}\right) -\frac{1}{2}{\rm tr}\left\{R+iB-iW\right\}\right] \left\langle{ \vec{u}\left|\right.}\zeta' \right\rangle. \end{multline}

We now show that eq.~\eqref{Eq:NdotAbs} results from the form of $\left|{\mathcal N}\right|$ as given in eq.~\eqref{Eq:fNAbs}. This follows by use of trace properties and the definition of $W$ from eq.~\eqref{Eq:TransformedHfinal} in the following identity,  
\[ \partial_t \det{A}={\rm tr}\left\{\left(\det{A}\right)A^{-1}\partial_t A\right\},\]
to get
\be \frac{\left|\dot{\mathcal N}\right|}{\left|{\mathcal N}\right|} = -\frac{1}{4}\frac{\det\left(UU^{\dag}\right)^{-5/4}}{\det\left(UU^{\dag}\right)^{-1/4}}\partial_t\det\left(UU^{\dag}\right) = -\frac{1}{4}{\rm tr}\left\{ U^{-1}\dot{U}+U^{-\dag}\dot{U^{\dag}}\right\} =-\frac{1}{2}{\mathfrak Re}\left\{{\rm tr}W\right\}.\ee

The solution of eq.~\eqref{Eq:NArgdot} is given by
\begin{multline} \partial_t\left(-\frac{1}{2}\arg\det U\right)= \partial_t\left(-\frac{1}{2}\mathfrak{Im}\log\det U\right)=\\= -\frac{1}{2}\mathfrak{Im}\frac{\partial_t\det U}{\det U}=-\frac{1}{2}\mathfrak{Im}\frac{{\rm tr}\left\{\left(\det U\right) U^{-1}\partial_tU\right\}}{\det U}=-\frac{1}{2}\mathfrak{Im}\left\{{\rm tr}W\right\}\end{multline}

We now consider the integral in eq.~\eqref{Eq:deltann}. Using eq.~\eqref{Eq:IntegrationIdentities} we get
\begin{multline} \delta_{\vec{n},\vec{n}'} = \left|\mathcal{N}\right|^2 {c}_{\vec{n}}^* {c}_{\vec{n}'}\int d^f\vec{u} \exp\left\{-\frac{1}{2} \vec{u}^t U^{-\dag}U^{-1}\vec{u}\right\}  \left(H_{\vec{n}}^{C}\left(U^{-*}\vec{u}\right)\right)^* H_{\vec{n}'}^C\left(U^{-*}\vec{u}\right)e^{i\sum_j \left(n_j-n_j'\right)\beta_jt} .\end{multline} 
Changing integration variables by $\vec{x}=U^{-*}\vec{u}$ and using the fact that $U^{-1}U^{*}=U^{\dag}U^{-t}$ since the latter is symmetric, we get 
\be \delta_{\vec{n},\vec{n}'} = \left|\mathcal{N}\right|^2 {c}_{\vec{n}}^* {c}_{\vec{n}'}\left|\det U^{*}\right|e^{i\sum_j \left(n_j-n_j'\right)\beta_jt} \mathcal{I}_{\vec{n},\vec{n}'},\ee 
where $\mathcal{I}_{\vec{n},\vec{n}'}$ is the integral 
\begin{multline} \mathcal{I}_{\vec{n},\vec{n}'}=\int d^f\vec{x} \exp\left\{-\frac{1}{2} \vec{x}^t C\vec{x}\right\} \left(H_{\vec{n}}^{C}\left(\vec{x}\right)\right)^* H_{\vec{n}'}^C\left(\vec{x}\right) = \delta_{\vec{n},\vec{n}'}{n_1}!\cdots {n_f}!\left(2\pi\right)^{f/2}\left(\det C\right)^{-1/2}, \end{multline} 
and the last equality follows from eq.~12.9(1) of \cite{Bateman}, using the fact that $C^*=C^{-1}$, which identifies the dual polynomials of $H_n\left(\vec{x}\right)$ (denoted $G_n\left(\vec{x}\right)$ in \cite{Bateman}), with their complex conjugates, $H_n\left(\vec{x}\right)^*$. 
Finally, since $\det C=\det U^{\dag} /\det U$, and using eq.~\eqref{Eq:fNAbs}, all time-dependent terms cancel and eq.~\eqref{Eq:N_n} results.

\end{appendix}
 
\bibliographystyle{../hunsrt}%abbrv

\bibliography{../bibfile}

\end{document}